# Thermal Conductivity due to Spins in the Two-Dimensional Spin System $Ba_2Cu_3O_4Cl_2$

Masumi Ohno[1], Takayuki Kawamata[1], Megumi Akoshima[2], and Yoji Koike[1]

[1]*Department of Applied Physics, Tohoku University, Sendai 980-8579, Japan*
[2]*National Institute of Advanced Industrial Science and Technology (AIST), Tsukuba, Ibaraki 305-8501, Japan*

We have measured the temperature dependences of the thermal conductivity of $Ba_2Cu_{3-x}M_xO_4Cl_2$ ($M$ = Pd, Ni, Co; $x$ = 0, 0.03) single crystals including two-dimensional (2D) $Cu_3O_4$ planes consisting of a strong 2D spin network of $Cu_A^{2+}$ spins and a weak 2D spin network of $Cu_B^{2+}$ spins. It has been found that the thermal conductivity due to spins, $\kappa_{spin}$, exists in the thermal conductivity parallel to the $Cu_3O_4$ plane owing to the strong 2D spin network of $Cu_A^{2+}$ spins and exhibits a broad peak around room temperature. The maximum value of $\kappa_{spin}$ is ~7 W/Km and comparable with that in $Nd_2CuO_4$ with almost the same 2D spin network of $Cu^{2+}$ spins. The $\kappa_{spin}$ has been found to be suppressed by 1% impurities on account the decrease in the mean free path of magnetic excitations, suggesting that $\kappa_{spin}$ is expected to be enhanced in 2D quantum spin systems such as $Ba_2Cu_3O_4Cl_2$ by reducing the amount of impurities in a single crystal. Moreover, it has concluded that the frustration between $Cu_A^{2+}$ and $Cu_B^{2+}$ spins little affects the existence of $\kappa_{spin}$.

## 1. Introduction

In several low-dimensional quantum spin systems, the thermal conductivity due to magnetic excitations, namely, due to spins, $\kappa_{spin}$, has been observed conspicuously. As for one-dimensional (1D) quantum spin systems, $\kappa_{spin}$ has been observed in antiferromagnetic (AF) spin systems of $Sr_2CuO_3$,[1–4] $SrCuO_2$,[2,5–7] $Sr_{14}Cu_{24}O_{41}$,[8–13] $Sr_2V_3O_9$,[14] and $ACoX_3$ ($A$ = Cs, Rb; $X$ = Cl, Br)[15] with the spin quantum number $S = 1/2$ and of $Y_2BaNiO_5$[16,17] and $AgVP_2S_6$[18] with $S = 1$. $\kappa_{spin}$ is especially large in $Sr_2CuO_3$, $SrCuO_2$, and $Sr_{14}Cu_{24}O_{41}$, which is understood to be owing to the large velocity of magnetic excitations, $v_{spin}$, due to the large AF superexchange interaction between the nearest neighbor $Cu^{2+}$ spins.[19] In two-dimensional (2D) quantum spin systems, $\kappa_{spin}$ has also been observed in AF spin systems of $La_2CuO_4$,[20–23] $Nd_2CuO_4$,[24] and $Sr_2CuO_2Cl_2$[25] with $S = 1/2$ and of $La_2NiO_4$ with $S = 1$.[22] $\kappa_{spin}$ is especially large in $La_2CuO_4$, $Nd_2CuO_4$, and $Sr_2CuO_2Cl_2$, which is understood to be due to the AF superexchange interaction as large as ~1500 K between the nearest neighbor $Cu^{2+}$ spins in the 2D spin network of $Cu^{2+}$ spins in the 2D $CuO_2$ plane. It is characteristic of these 2D quantum spin systems that $\kappa_{spin}$ exhibits the maximum around room temperature, which is suitable for the application of the large $\kappa_{spin}$.

The compound $Ba_2Cu_3O_4Cl_2$ is regarded as a 2D quantum spin system. The crystal structure is layered and composed of an alternate stack of the 2D $Cu_3O_4$ plane and the $Ba_2Cl_2$ blocking layer along the $c$-axis, as shown in Fig. 1(a). There are two Cu sites, namely, $Cu_A$ and $Cu_B$ sites in the $Cu_3O_4$ plane, where $Cu_A^{2+}$ ions with $S = 1/2$ form the abovementioned 2D $CuO_2$ plane and $Cu_B^{2+}$ ions with $S = 1/2$ are alternately located at the center of the square composed of four $Cu_A^{2+}$ ions, as shown in Fig. 1(b). The Raman scattering experiment has revealed that the exchange interaction between the nearest neighbor $Cu_A^{2+}$ spins, $J_A$, in the 2D spin network of $Cu_A^{2+}$ spins is AF and ~1500 K, while that between the nearest neighbor $Cu_B^{2+}$ spins, $J_B$, in the 2D spin network of $Cu_B^{2+}$ spins is AF and ~120 K.[26] The inelastic neutron scattering experiment has revealed that the exchange interaction between the adjacent $Cu_A^{2+}$ and $Cu_B^{2+}$ spins, $J_{AB}$, is ferromagnetic and about -140 K.[27] Therefore, frustration exists between $Cu_A^{2+}$ and $Cu_B^{2+}$ spins. From the temperature dependence of the magnetization, it has been found that two magnetic phase transitions occur at $T_{N1}$ ~ 30 K and at $T_{N2}$ ~ 330 K.[28–31] The elastic neutron scattering experiment has revealed that $T_{N1}$ and $T_{N2}$ are AF transition temperatures of $Cu_B^{2+}$ and $Cu_A^{2+}$ spins, respectively.[31]

Since $Ba_2Cu_3O_4Cl_2$ includes the 2D $CuO_2$ plane as well as $La_2CuO_4$, $Nd_2CuO_4$, and $Sr_2CuO_2Cl_2$, large $\kappa_{spin}$ is expected to be observed. However, the frustration between $Cu_A^{2+}$ and $Cu_B^{2+}$ spins in the $Cu_3O_4$ plane may interrupt the appearance of large $\kappa_{spin}$. In this paper, we have grown large-sized single crystals of $Ba_2Cu_{3-x}M_xO_4Cl_2$ ($M$ = Pd, Ni, Co; $x$ = 0, 0.03) including no and 1% impurities of nonmagnetic $Pd^{2+}$ ($S$ = 0) and magnetic $Ni^{2+}$ ($S$ = 1) and $Co^{2+}$ ($S$ = 3/2) and measured the thermal conductivity, to investigate whether $\kappa_{spin}$ exists in $Ba_2Cu_3O_4Cl_2$. Moreover, the origin of the maximum of $\kappa_{spin}$ around room temperature in the 2D quantum spin systems consisting of the 2D $CuO_2$ plane has been discussed.

## 2. Experimental

Single crystals of $Ba_2Cu_{3-x}M_xO_4Cl_2$ ($M$ = Pd, Ni, Co; $x$ = 0, 0.03) were grown by the floating-zone method without the use of solvent, referring to the literature by Yamada *et al.*[32] The growth was performed under flowing $O_2$ of 4 atm at the rate of 10 mm/h. The grown crystals were confirmed by the powder x-ray diffraction to be of the single phase of $Ba_2Cu_3O_4Cl_2$ without any impurity phases. The diffraction spots of the x-ray back-Laue photography were very sharp, indicating the good quality of the obtained single crystals. The chemical composition was confirmed by the inductively coupled plasma mass spectrometry (ICP-MS) to coincide with the nominal composition, as listed in Table I. For the characterization of the obtained single crystals, the magnetic susceptibility was also measured using a SQUID magnetometer (Quantum Design, MPMS).

Thermal conductivity measurements were carried out by the conventional four-terminal steady-state method. On the thermal conductivity data obtained by this method, the effect of the thermal radiation is marked at high temperatures above ~150 K. Therefore, the thermal conductivity due to the thermal radiation, $\kappa_{radiation}$, has to be subtracted from the observed thermal conductivity. The temperature dependence of $\kappa_{radiation}$ was estimated as follows. First, the thermal conductivity was measured at room temperature by the laser-flash method giving a rather correct value with little influence of the thermal radiation.[33] Next, the value of $\kappa_{radiation}$ at room temperature was estimated as the difference between the values obtained by these two methods. Finally, the temperature dependence of $\kappa_{radiation}$ was estimated using the equation, $\kappa_{radiation} = 4\sigma\varepsilon A_s T^3$,[34] where $\sigma$ is the Stefan-Boltzmann constant, $\varepsilon$ the reflectivity, $A_s$ the surface area, and the $T$ temperature of a sample. The value of $\varepsilon A_s$ was calculated from the value of $\kappa_{radiation}$ at room temperature. All the thermal conductivity data shown in this paper were obtained after the subtraction of $\kappa_{radiation}$. The value of $\kappa_{radiation}$ at room temperature was 30 – 50 % of that of the thermal conductivity measured by the four-terminal steady-state method.

For the estimate of the Debye temperature, the specific heat was measured by the thermal relaxation method using a physical property measurement system (Quantum Design, PPMS).

## 3. Results and discussion

Figure 2 shows the temperature dependences of the magnetization in a magnetic field of 0.5 T applied in the *ab*-plane, namely, in the $Cu_3O_4$ plane of $Ba_2Cu_{3-x}M_xO_4Cl_2$ ($M$ = Pd, Ni, Co; $x$ = 0, 0.03) on zero-field cooling and field cooling. For $x$ = 0, it is found that, with decreasing temperature, the magnetization increases a little at high temperatures below 400 K, drastically jumps up at $T_{N2}$ = 330 K, gradually increases, exhibits a broad peak around 80 K, and decreases. Then, the magnetization shows a hysteresis at low temperatures below $T_{N1}$ = 32 K. This temperature dependence of the magnetization for $x$ = 0 is almost the same as that in the former report by Noro *et al*.[29] For $x$ = 0.03 of $M$ = Pd and Ni, $T_{N2}$ is clearly found to decrease compared with those of $x$ = 0, while $T_{N1}$ increases for $x$ = 0.03 of $M$ = Co. These changes in $T_{N2}$ by Pd, Ni, and Co impurities indicate that these impurities are substituted for Cu, though the microscopic origin of the increase or decrease in $T_{N2}$ is not clear in detail.

Figure 3(a) shows the temperature dependences of the thermal conductivity along the [110] direction, namely, along the $Cu_A$-O-$Cu_A$ direction in the $Cu_3O_4$ plane, $\kappa_{[110]}$, and along the [001] direction perpendicular to the $Cu_3O_4$ plane, $\kappa_{[001]}$, of $Ba_2Cu_3O_4Cl_2$. It is found that there is a large anisotropy between $\kappa_{[110]}$ and $\kappa_{[001]}$. That is, $\kappa_{[001]}$ exhibits a peak at ~30 K and monotonously decreases with increasing temperature. On the other hand, $\kappa_{[110]}$ exhibits a peak at ~25 K and decreases with increasing temperature, but the decrease becomes gentle at high temperatures above ~80 K and $\kappa_{[110]}$ shows a broad peak around room temperature. Since the [110] direction is parallel to the $Cu_3O_4$ plane including 2D $CuO_2$ plane and this kind of behavior of the thermal conductivity has been observed in $La_2CuO_4$ also,[20] the peaks at low temperatures of ~30 K and ~25 K in $\kappa_{[001]}$ and $\kappa_{[110]}$, respectively, are inferred to be caused by the contribution of the thermal conductivity due to phonons, $\kappa_{phonon}$, while the broad peak around room temperature only in $\kappa_{[110]}$ is inferred to be due to $\kappa_{spin}$.

To estimate $\kappa_{spin}$, at first the estimate of $\kappa_{phonon}$ is necessary, for the observed thermal conductivity is regarded as the sum of $\kappa_{phonon}$ and $\kappa_{spin}$. In $Ba_2Cu_3O_4Cl_2$, it is unsuitable to estimate $\kappa_{spin}$ simply as the difference between $\kappa_{[110]}$ and a constant multiple of $\kappa_{[001]}$, assuming that $\kappa_{[001]}$ is due to only $\kappa_{phonon}$. This is because $\kappa_{phonon}$ is very anisotropic, considering the large differences between $\kappa_{[110]}$ and $\kappa_{[001]}$ of the magnitude and broadening of the low-temperature peak due to $\kappa_{phonon}$. Therefore, the contributions of $\kappa_{phonon}$ to both $\kappa_{[110]}$ and $\kappa_{[001]}$ have to be estimated using a suitable theoretical model. The $\kappa_{phonon}$ is simply given by the following equation based on the Debye model.[35]

$$\kappa_{phonon} = \frac{k_B}{2\pi v_{phonon}} \left(\frac{k_B}{\hbar}\right)^3 T^3 \int_0^{\Theta_D/T} \frac{x^4 e^x}{(e^x-1)^2} \tau(x,T) dx, \quad (1)$$

where $x = \hbar\omega/k_B T$, $\omega$ is the phonon angular frequency, $\hbar$ the Planck constant, $k_B$ the Boltzmann constant, $\Theta_D$ the Debye temperature, $v_{phonon}$ the phonon velocity, and $\tau(x, T)$ the relaxation time of the phonon scattering. The $v_{phonon}$ is described as

$$v_{phonon} = \Theta_D \left(\frac{k_B}{\hbar}\right) (6\pi^2 n)^{-1/3}, \quad (2)$$

where $n$ is the number density of atoms. The phonon scattering rate $\tau^{-1}(x, T)$ is given by the sum of scattering rates due to several scattering processes as follows,

$$\tau^{-1}(x,T) = \tau^{-1}(\omega,T) = \frac{v_{phonon}}{L_b} + A\omega^4 + D\omega + B\omega^2 T \exp\left(-\frac{\Theta_D}{bT}\right), \quad (3)$$

where the first term represents the phonon scattering by boundaries; the second, the phonon scattering by point defects; the third, the phonon scattering by lattice distortions; the fourth, the phonon-phonon scattering in the umklapp process. $L_b$ is the distance between two temperature terminals and $A$, $D$, $B$, and $b$ are fitting parameters. Using Eqs. (1) – (3) and putting $\Theta_D$ at 470 K from the specific heat measurements, $\kappa_{phonon}$ was estimated as drawn by dashed lines in Fig. 3(a). It is found that $\kappa_{[001]}$ is well fitted in the whole temperature region, indicating that $\kappa_{[001]}$ is due to only $\kappa_{phonon}$. This is reasonable, because $\kappa_{spin}$ is not expected to be observed in $\kappa_{[001]}$ on account of the very weak magnetic correlation along the [001] direction. In the case of the estimate of $\kappa_{phonon}$ in $\kappa_{[110]}$, only the data of $\kappa_{[110]}$ at low temperatures below the peak temperature of ~25 K were used for the fit with Eqs. (1) – (3), assuming that there was little $\kappa_{spin}$ at the low temperatures. This assumption is reasonable, because $\kappa_{spin}$ is expected to decrease in proportion to $T^2$ with decreasing temperature at low temperatures due to the $T^2$ dependence of the specific heat in a 2D AF spin system.[36] In this case, $\kappa_{phonon}$ is a little overestimated, while $\kappa_{spin}$ is a little underestimated. However, the estimate of large $\kappa_{spin}$ around room temperature is little influenced. Values of the parameters used for the best fit are listed in Table II. It is found that values of $A$ and $D$ in $\kappa_{[001]}$ are much larger than those in $\kappa_{[110]}$, indicating the large phonon scattering rate by point defects and lattice distortions probably in the [001] direction. This is reasonable, because the atomic bonding in the [001] direction is much weaker than in the [110] direction, as suggested by the strong cleavability in the (001) plane. The value of $B$ in $\kappa_{[001]}$ is larger than that in $\kappa_{[110]}$, indicating the large phonon-phonon scattering rate in the umklapp process in the [001] direction. This is also reasonable, taking into account the fact that the size of the first Brillouin zone in the [001] direction is smaller than that in the [110] direction owing to the $c$-axis length being larger than the $a$- and $b$-axis lengths. In addition, these values of the parameters listed in Table II are comparable with those obtained in several low-dimensional quantum spin systems,[2,4,5,14,15] except for $D$ values. The large $D$ values in $Ba_2Cu_3O_4Cl_2$ are inferred to be due to the strong cleavability. Accordingly, the estimate of $\kappa_{phonon}$ in $\kappa_{[110]}$ seems to be appropriate. Thereupon, looking at $\kappa_{[110]}$ shown in Fig. 3(a), it is clearly found that another large contribution to the thermal conductivity except for $\kappa_{phonon}$ exists at high temperatures above ~80 K, so that the large contribution is regarded as being due to $\kappa_{spin}$.

To confirm the large contribution of $\kappa_{spin}$ at high temperatures above ~80 K, the temperature dependences of the thermal conductivity along the [110] direction have been measured for $Ba_2Cu_{3-x}M_xO_4Cl_2$ ($M$ = Pd, Ni, Co; $x$ = 0.03) including 1% impurities of Pd, Ni, and Co, as shown in Figs. 3(b), 3(c), and 3(d), respectively. It is found that both the peak at ~25 K and the broad peak around room temperature are suppressed by the 1% impurities, though the peak at ~25 K in the crystal with 1% Co is not suppressed so much.

To estimate $\kappa_{spin}$ in $Ba_2Cu_{3-x}M_xO_4Cl_2$ ($M$ = Pd, Ni, Co; $x$ = 0.03), $\kappa_{phonon}$ was also estimated using Eqs. (1) – (3) as in the above case of $x$ = 0. The best fit result of $\kappa_{phonon}$ is shown by dashed lines in Figs. 3(b), 3(c), and 3(d), and values of the used parameters are listed in Table II. It is found that the value of $B$ relating to the phonon-phonon scattering is not so dependent on the kind of impurity, which is reasonable. On the other hand, values of $A$ and $D$ are dependent on the kind of impurity. Since they are also dependent on the crystalline quality, the difference of the $A$ and $D$ values among 1% impurity-substituted crystals is not discussed clearly. However, it is clearly found that another contribution to the thermal conductivity except for $\kappa_{phonon}$, namely, $\kappa_{spin}$ still exists at high temperatures above ~80 K in $Ba_2Cu_{3-x}M_xO_4Cl_2$ ($M$ = Pd, Ni, Co; $x$ = 0.03).

Figure 4 displays the temperature dependences of $\kappa_{spin}$ obtained by subtracting $\kappa_{phonon}$ from the observed $\kappa_{[110]}$ for $Ba_2Cu_{3-x}M_xO_4Cl_2$ ($M$ = Pd, Ni, Co; $x$ = 0, 0.03). It is found that $\kappa_{spin}$ of $x$ = 0 clearly exhibits a peak at ~310 K and that this peak is suppressed by the 1% impurities and most suppressed by Pd. Since $S$ values of $Pd^{2+}$, $Ni^{2+}$, and $Co^{2+}$ are 0, 1, and 3/2, respectively, and different from $S$ = 1/2 of $Cu^{2+}$, this is interpreted as being due to the strong scattering of magnetic excitations by ions with different $S$ values. It appears that especially nonmagnetic $Pd^{2+}$ ions with $S = 0$ strongly scatter magnetic excitations, leading to the strongest suppression of $\kappa_{spin}$. Such impurity effects on $\kappa_{spin}$ have been also observed in $La_2CuO_4$.[21,23] Therefore, the present results strongly support the existence of $\kappa_{spin}$ in $Ba_2Cu_3O_4Cl_2$ and the maximum value of $\kappa_{spin}$ is ~7 W/Km. This is comparable with that in $Nd_2CuO_4$.[24] Since the value of $J_A$ is comparable with the AF superexchange interaction between the nearest neighbor $Cu^{2+}$ spins in the $CuO_2$ plane of $Nd_2CuO_4$,[37] it is understood that the 2D spin network consisting of $Cu_A^{2+}$ spins in the $Cu_3O_4$ plane contribute to $\kappa_{spin}$. Accordingly, it follows that the frustration between $Cu_A^{2+}$ and $Cu_B^{2+}$ spins little affects the existence of $\kappa_{spin}$.

Here, we estimate the mean free path of magnetic excitations, $l_{spin}$. In general, $\kappa_{spin}$ is given by the following equation,

$$\kappa_{spin} = \Sigma_k C_k v_k l_k = \frac{1}{(2\pi)^d} \int C_k v_k l_k dk , \qquad (4)$$

where $C_k$, $v_k$, and $l_k$ are the specific heat, velocity, and mean free path of the magnetic excitation with the wave number $k$, respectively. $d$ is the dimension of a spin net network. Assuming that both $v_k$ and $l_k$ are independent of $k$ and $T \ll J/k_B$ ($J$: the exchange interaction between the nearest neighbor spins), eq. (4) is transformed as

$$\kappa_{spin} = \frac{k_B^3 l_{spin}}{2\pi c \hbar^2 v_{spin}} T^2 \int_0^{x_{max}} \frac{x^3 e^x}{(e^x - 1)^2} dx, \qquad (5)$$

where $c$ is the $c$-axis length, $x = \hbar v_{spin}/k_B T$, $x_{max} = 2\sqrt{\pi}\hbar v_{spin}/a k_B T$, and $a$ is the $a$-axis length. Both $l_{spin}$ and $v_{spin}$ are assumed to be independent of temperature and $v_{spin}$ is given by the following equation,[25]

$$v_{spin} = \sqrt{8} S Z_c J a / \hbar, \qquad (6)$$

where $Z_c$ is 1.18 and called the Oguchi correction.[38] $\kappa_{spin}$ is roughly fitted using Eqs. (5) and (6), as shown by dashed lines in Fig. 4. Values of $l_{spin}$ used for the best fit, which are regarded as the upper limit values due to the scattering of magnetic excitations by impurities at low temperatures of $T \ll J/k_B$, are listed in Table III. It is found that the value of $l_{spin}$ of $x = 0$ decreases by the 1% impurities and markedly decreases by Pd. The marked decrease in $l_{spin}$ by Pd is understood to be due to the division of the spin network by nonmagnetic $Pd^{2+}$ ions. On the other hand, the decrease in $l_{spin}$ by Ni and Co is not so marked, because magnetic $Ni^{2+}$ and $Co^{2+}$ ions do not divide the spin network completely. The decrease in $l_{spin}$ by the 1% impurities indicates that $l_{spin}$ already reaches the upper limit at ~310 K. That is, the scattering of magnetic excitations by impurities is dominant at low temperatures below ~310 K. As in the case of 1D quantum spin systems such as $Sr_2CuO_3$[1-4] and $SrCuO_2$,[2,5-7] accordingly, it is possible to enhance $\kappa_{spin}$ in 2D quantum spin systems such as $Ba_2Cu_3O_4Cl_2$ by reducing the amount of impurities in a single crystal.

Finally, it may be worthwhile pointing out the possibility of the existence of $\kappa_{spin}$ due to the spin network of $Cu_B^{2+}$ spins. Taking into account the fact that $\kappa_{spin}$ due to the 2D spin network of $Cu_A^{2+}$ spins (with $J_A \sim 1500$ K and $T_{N2} \sim 330$ K) exhibits a peak at ~310 K, it is possible that $\kappa_{spin}$ due to the 2D spin network of $Cu_B^{2+}$ spins (with $J_B \sim 120$ K and $T_{N1} \sim 30$ K) exists and exhibits a peak at a low temperature below $T_{N1}$.[26,28-31] As shown in Fig. 4, in fact, $\kappa_{spin}$ increases with decreasing temperature at low temperatures below ~25 K. This is caused by the misfit of $\kappa_{[110]}$ with $\kappa_{phonon}$ based on the Debye model at low temperatures below ~25 K. Therefore, it is possible that $\kappa_{spin}$ due to the 2D spin network of $Cu_B^{2+}$ spins exists at low temperatures, though it is very hard to separate $\kappa_{spin}$ from $\kappa_{phonon}$.

## 4. Summary

We have grown large-sized single crystals of $Ba_2Cu_{3-x}M_xO_4Cl_2$ ($M$ = Pd, Ni, Co; $x$ = 0, 0.03) and measured the thermal conductivity. For $x$ = 0, $\kappa_{[001]}$ perpendicular to the $Cu_3O_4$ plane has been found to exhibit a peak at ~30 K and monotonously decreases with increasing temperature, while $\kappa_{[110]}$ parallel to the $Cu_3O_4$ plane has been found to exhibit a peak at ~25 K and a broad peak around room temperature. It has been concluded that the peaks at low temperatures of ~30 K and ~25 K in $\kappa_{[001]}$ and $\kappa_{[110]}$, respectively, are caused by the contribution of $\kappa_{phonon}$, because the peaks and the whole temperature dependence of $\kappa_{[001]}$ were well fitted with $\kappa_{phonon}$ based on the Debye model. On the other hand, the broad peak around room temperature in $\kappa_{[110]}$ has been concluded to be due to $\kappa_{spin}$, because $\kappa_{[110]}$ is parallel to the $Cu_3O_4$ plane including the 2D spin network of $Cu_A^{2+}$ spins with $J_A$ as large as ~ 1500 K and because the maximum value of $\kappa_{spin}$ of ~7 W/Km estimated is comparable with that in $Nd_2CuO_4$ with almost the same 2D spin network of $Cu^{2+}$ spins. It has also supported the existence of $\kappa_{spin}$ that the $\kappa_{spin}$ was suppressed by 1% impurities of magnetic $Ni^{2+}$ and $Co^{2+}$ with $S$ values different from that of $Cu^{2+}$ and most suppressed by 1% impurities of nonmagnetic $Pd^{2+}$. Accordingly, it has concluded that the frustration between $Cu_A^{2+}$ and $Cu_B^{2+}$ spins little affects the existence of $\kappa_{spin}$. Moreover, it has been found that the suppression of $\kappa_{spin}$ by 1% impurities is due to the decrease in $l_{spin}$. Therefore, it is expected to enhance $\kappa_{spin}$ in 2D quantum spin systems such as $Ba_2Cu_3O_4Cl_2$ by reducing the amount of impurities in a single crystal.

In addition, it has been found that $\kappa_{\rm spin}$ due to the 2D spin network of $Cu_B^{2+}$ spins with $J_B$ as small as ~ 120 K may exist and exhibit a peak at a low temperature below $T_{N1}$.

**Acknowledgments**

This work was supported by JSPS KAKENHI Grant Numbers 16K06716. We would like to thank Y. Nakano of Technical Division, School of Engineering, Tohoku University, for her aid in the ICP-MS analysis. Figure 1 was drawn using VESTA.[39]

Table I. Chemical compositions obtained by the inductively coupled plasma mass spectrometry (ICP-MS) for $Ba_2Cu_{3-x}M_xO_4Cl_2$ ($M$ = Pd, Ni, Co; $x$ = 0, 0.03).

| Sample | Ba | Cu | M |
|---|---|---|---|
| $x = 0$ | 1.990 | 3.000 | |
| $x(Pd) = 0.03$ | 2.081 | 2.973 | 0.027 |
| $x(Ni) = 0.03$ | 1.925 | 2.971 | 0.029 |
| $x(Co) = 0.03$ | 1.969 | 2.972 | 0.028 |

Table II. Parameters used for the best fit of the temperature dependence of the thermal conductivity $\kappa_{[001]}$ and $\kappa_{[110]}$ in $Ba_2Cu_{3-x}M_xO_4Cl_2$ ($M$ = Pd, Ni, Co; $x$ = 0, 0.03) with Eqs. (1) – (3).

| Sample | Direction | $L_b$ ($10^{-3}$ m) | $A$ ($10^{-42}$ s$^3$) | $D$ ($10^{-3}$) | $B$ ($10^{-17}$ s/K) | $b$ |
|---|---|---|---|---|---|---|
| $x = 0$ | [001] | 0.700 | 1.90 | 1.20 | 1.83 | 7.2 |
| | [110] | 1.06 | 0.140 | 0.127 | 0.620 | 6.9 |
| $x$(Pd) = 0.03 | [110] | 0.750 | 0.770 | 0.190 | 0.450 | 7.3 |
| $x$(Ni) = 0.03 | [110] | 0.588 | 0.570 | 0.120 | 0.620 | 6.9 |
| $x$(Co) = 0.03 | [110] | 0.888 | 0.140 | 0.127 | 0.580 | 5.9 |

Table III. Mean free path $l_{spin}$ used for the best fit of $\kappa_{spin}$ in $Ba_2Cu_{3-x}M_xO_4Cl_2$ ($M$ = Pd, Ni, Co; $x$ = 0, 0.03) with Eqs. (5) and (6).

| Sample | $l_{spin}$ (Å) |
|---|---|
| $x = 0$ | 425±25 |
| $x$(Pd) = 0.03 | 70±15 |
| $x$(Ni) = 0.03 | 220±20 |
| $x$(Co) = 0.03 | 140±20 |

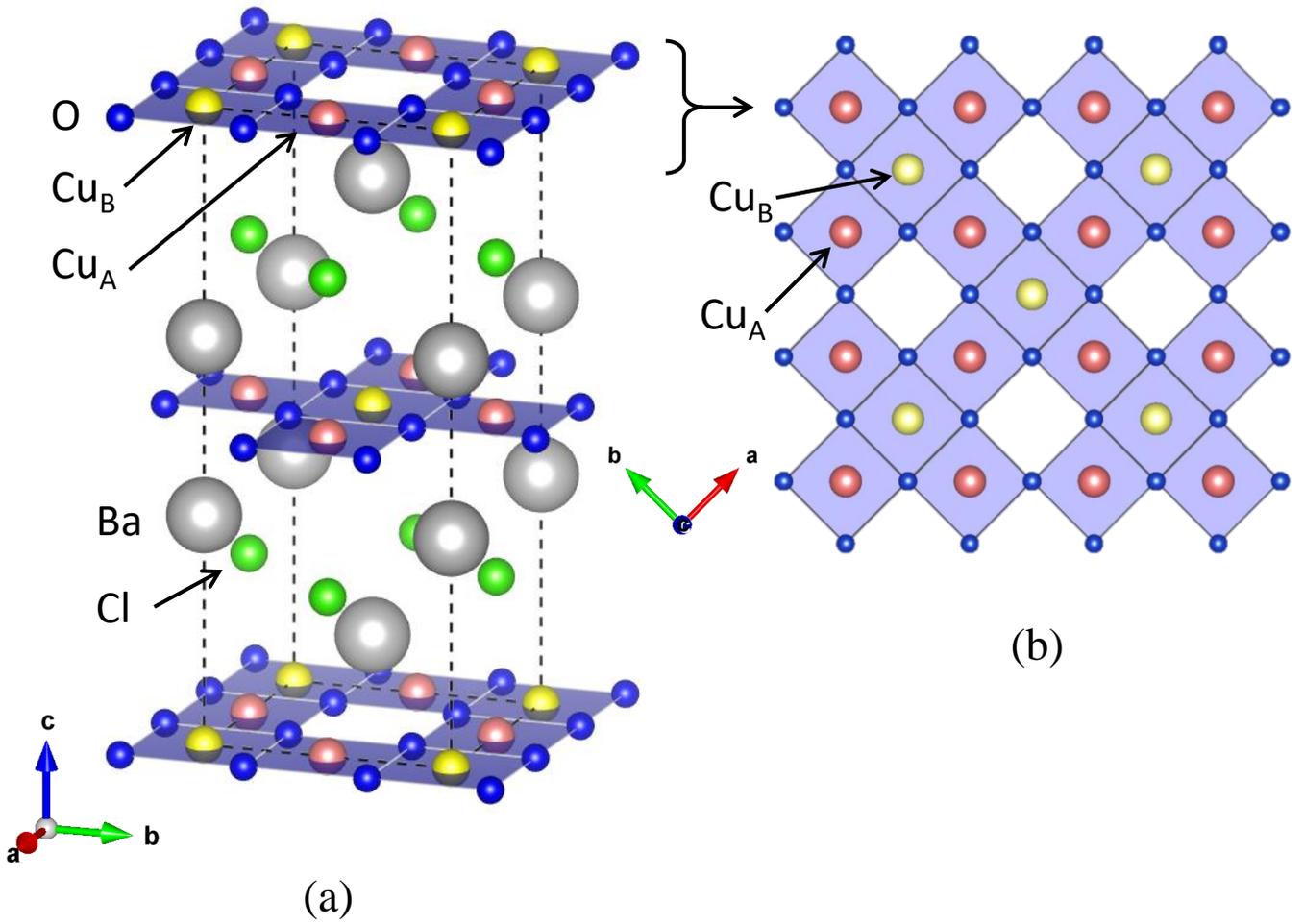

Fig. 1. (color online) (a) Crystal structure of $Ba_2Cu_3O_4Cl_2$. The dashed lines indicate the unit cell. (b) Schematic picture of the $Cu_3O_4$ plane with two Cu sites, namely, $Cu_A$ sites (red spheres) and $Cu_B$ sites (yellow spheres).

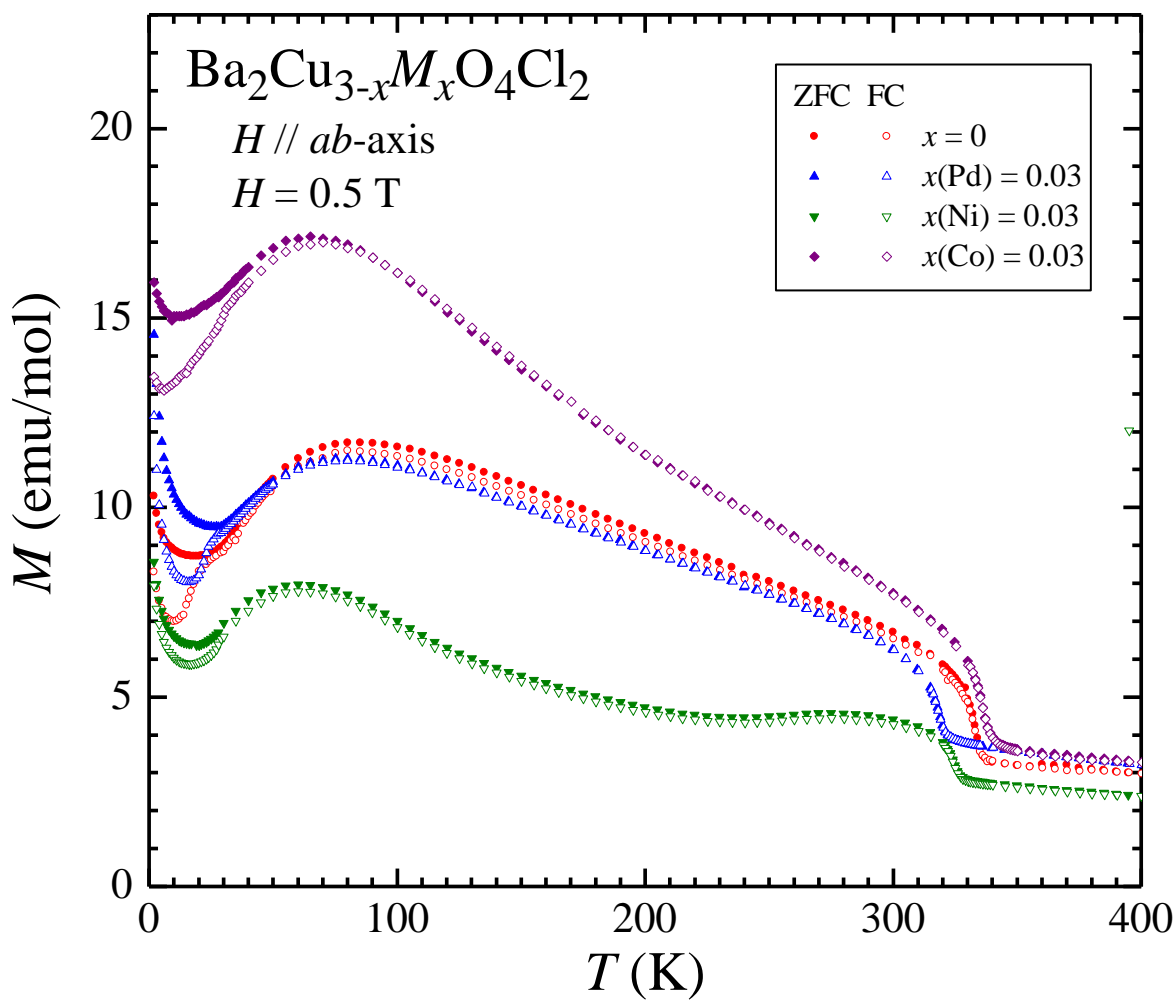

Fig. 2. (color online) Temperature dependences of the magnetization in a magnetic field of 0.5 T applied in the *ab*-plane of $Ba_2Cu_{3-x}M_xO_4Cl_2$ ($M$ = Pd, Ni, Co; $x$ = 0, 0.03) on zero-field cooling (open circles) and on field cooling (closed circles).

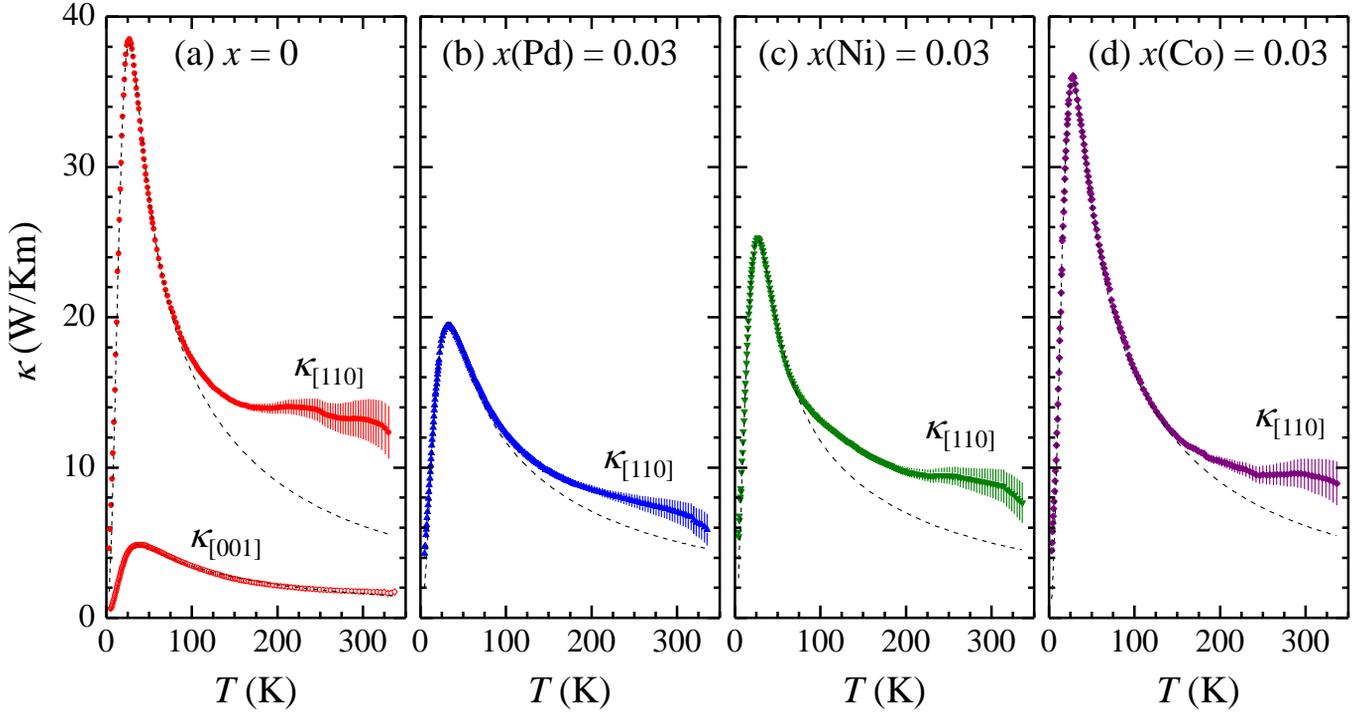

Fig. 3. (color online) Temperature dependences of the thermal conductivity along the [110] direction, namely, along the $Cu_A$-O-$Cu_A$ direction in the $Cu_3O_4$ plane, $\kappa_{[110]}$, of $Ba_2Cu_{3-x}M_xO_4Cl_2$ with (a) $x = 0$, (b) $M = $ Pd and $x = 0.03$, (c) $M = $ Ni and $x = 0.03$, and (d) $M = $ Co, $x = 0.03$. For $x = 0$ in (a), the temperature dependence of the thermal conductivity along the [001] direction perpendicular to the $Cu_3O_4$ plane, $\kappa_{[001]}$, is also displayed. Error bars are due to errors in the thermal conductivity measurements performed by the laser-flash method at room temperature for the estimate of $\kappa_{radiation}$. Dashed lines are $\kappa_{phonon}$ estimated using eqs. (1) – (3) based on the Debye model.

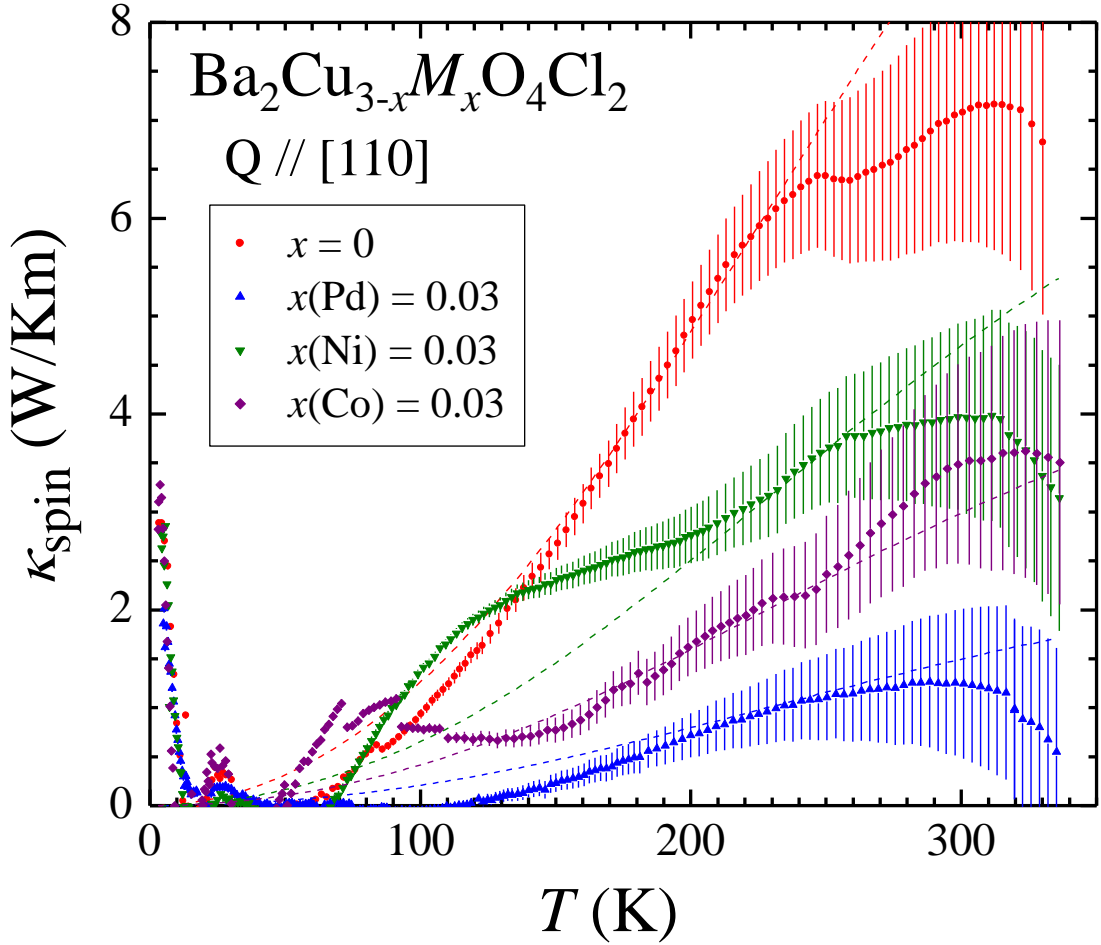

Fig. 4. (color online) Temperature dependences of $\kappa_{\mathrm{spin}}$ obtained by subtracting $\kappa_{\mathrm{phonon}}$ from the observed $\kappa_{[110]}$ for Ba$_2$Cu$_{3-x}$M$_x$O$_4$Cl$_2$ ($M$ = Pd, Ni, Co; $x$ = 0, 0.03). Error bars are due to errors in the thermal conductivity measurements performed by the laser-flash method at room temperature for the estimate of $\kappa_{\mathrm{radiation}}$. Dashed lines are the best fit results using Eqs. (5) and (6).